\documentclass[final]{aipproc}

\layoutstyle{6x9}
\usepackage{amsmath,amsthm,amsfonts}
\usepackage[utf8x]{inputenc}
\begin{document}

\title[Inclusion of matter in inhomogeneous loop quantum cosmology]{Inclusion of matter in inhomogeneous loop quantum cosmology}

\author{D. Martín-de~Blas}{address={Instituto de Estructura de la Materia, IEM-CSIC, Serrano 121, 28006 Madrid, Spain.}, email={daniel.martin@iem.cfmac.csic.es}}

\author{M. Martín-Benito}{address={MPI f\"ur Gravitational Physics, Albert Einstein Institute,
Am M\"uhlenberg 1, D-14476 Potsdam, Germany.}, email={mercedes@aei.mpg.de}}

\author{G.~A. Mena Marugán}{address={Instituto de Estructura de la Materia, IEM-CSIC, Serrano 121, 28006 Madrid, Spain.}, email={mena@iem.cfmac.csic.es}}



\begin{abstract}
We study the hybrid quantization of the linearly polarized Gowdy $T^3$ model with a massless scalar field with the same symmetries as the metric. For simplicity, we quantize its restriction to the model with \emph{local rotational symmetry}. Using this hybrid approach, the homogeneous  degrees of freedom of the geometry are quantized {\sl \`a la loop}, leading to the resolution of the cosmological singularity. A Fock quantization is employed both for the matter and the gravitational inhomogeneities. Owing to the inclusion of the massless scalar field this system allows us to modelize flat Friedmann-Robertson-Walker cosmologies filled with inhomogeneities propagating in one direction, providing a perfect scenario to study the quantum back-reaction of the inhomogeneities on the polymeric homogeneous and isotropic background.
\end{abstract}

\maketitle

\section{Introduction}
Loop Quantum Cosmology (LQC) \cite{lqc1,lqc2,lqc3} applies the techniques of Loop Quantum Gravity (LQG) \cite{lqg} to quantize symmetry reduced systems. In this way, several homogeneous cosmologies have been satisfactory quantized, obtaining new quantum gravity phenomena that lead
to the resolution of cosmological singularities \cite{aps}.

In order to extend LQC to inhomogeneous cosmological systems, a hybrid quantization has been developed and applied to the vacuum Gowdy $T^3$ model with linear polarization \cite{hybrid1,hybrid2,hybrid3,hybrid4}. This hybrid quantization assumes that the most relevant quantum gravity effects are encoded in the homogeneous degrees of freedom, and therefore only those are quantized by using the polymeric approach of LQC. In order to deal with the field complexity of the inhomogeneities, a Fock quantization is employed for them.

The study of more realistic models calls for the inclusion of matter. With this aim we introduce in the Gowdy $T^{3}$ model a minimally coupled massless scalar field with the same symmetries as the geometry. Owing to the inclusion of matter, the model admits homogeneous and isotropic solutions of the flat Friedmann-Robertson-Walker (FRW) type. Then, this system provides a suitable setting to analyze the quantum interactions between the inhomogeneities and a(n approximated) loop quantum FRW background. The hybrid quantization of this model was studied in \cite{hybrid5}. For simplicity, and without loss of generality, we consider here the submodel consisting of solutions with \emph{local rotational symmetry}  (LRS), that we will call LRS-Gowdy. The hybrid quantization leads to the resolution of the initial singularity and the recovering of the standard Fock quantization for the inhomogeneities \cite{cmmv,ccm}. Controlled approximations to the LRS-Gowdy model can be developed to obtain physically interesting 
solutions.

\section{Hybrid Quantization}

The Gowdy $T^{3}$ model describes globally hyperbolic spacetimes with the spatial topology of a three torus and with two axial and hypersurface orthogonal Killing vector fields, $\partial_\sigma$ and $\partial_\delta$. We choose coordinates $\{t,\theta,\sigma,\delta\}$ adapted to the symmetries, such that the only spatial dependence of our fields is on the coordinate $\theta \in S^1$. The Gowdy $T^{3}$ model is symmetric under the interchange of $\sigma$ and $\delta$, so that it has a subset of classical solutions with LRS, those in which the behavior in both directions $\sigma$ and $\delta$ is identical. We will focus on this LRS-Gowdy model. We perform a partial gauge fixing as in the vacuum case \cite{hybrid2}, obtaining a reduced phase space that can be split in two sectors by expanding the fields in Fourier modes. The \emph{homogeneous sector} contains the zero modes and is equivalent to the phase space of the LRS-Bianchi I model with a homogeneous massless scalar field, $\phi$. The nonzero modes of 
both gravitational and matter fields, $\xi$ and $\varphi$ respectively, form the \emph{inhomogeneous sector}. Two global constraints remain to be imposed at the quantum level: a momentum constraint $\mathcal C_{\theta}$ that generates translations in the circle and only involves the inhomogeneous sector, and the Hamiltonian constraint, $\mathcal{C}=\mathcal{C}_{\text{hom}}+\mathcal{C}_{\text{inh}}$, formed by a homogeneous term that corresponds to the Hamiltonian constraint of the LRS-Bianchi I model with a homogeneous scalar $\phi$, and by an additional term that couples homogeneous and inhomogeneous sectors.

For the homogeneous sector we employ the quantization of the Bianchi I model in LQC \cite{awe}, adapted to the LRS model. Then we start by describing it in terms of Ashtekar-Barbero variables of the LRS-Bianchi I model with three-torus topology. In a diagonal gauge these variables are given by the two independent components of the densitized triad, $p_{j}$, and of the $su(2)$-connection, $c_{j}$, with $j=\theta$ or $\perp$ (and $\perp$ denoting both $\sigma$ and $\delta)$. The nonvanishing Poisson brackets are $\{c_{\theta}, p_{\theta}\}=2\{c_{\perp}, p_{\perp}\}=8\pi \gamma G$, where $\gamma$ is the Immirzi parameter and $G$ is the Newton constant. The representation of these variables in LQC leads to a kinematical Hilbert space with a \emph{discrete inner product}, so that the triad operators $\hat{p}_{j}$ have a discrete spectrum equal to the real line.
Then, an orthonormal basis of the LRS-Bianchi I kinematical Hilbert space is given by the eigenstates of the two operators $\hat{p}_{j}$, denoted by $\{|p_\theta,p_\perp\rangle\ |\ p_{\theta}, p_{\perp}\in \mathbb{R}\}$. As a consequence of the discreteness of the representation, there is no well-defined operator corresponding to the connection coefficients $c_{j}$, but rather to their quasi-periodic functions, denoted by  $\mathcal{N}_{\mu_{j}}(c_{j})=e^{i c_{j}\mu_{j}/2}$, with $\mu_j\in\mathbb{R}$. The associated operators  $\hat{\mathcal{N}}_{\mu_{j}}$ act by translation on the states $|p_j\rangle$. The regularization of the Hamiltonian constraint, that classically depends on $c_j$, is attained by replacing the classical variable $c_j$ with the operator $(\hat{\mathcal{N}}_{\bar\mu_{j}}-\hat{\mathcal{N}}_{-\bar\mu_{j}})/(2i\bar{\mu}_{j})$, where $\bar{\mu}_{\theta}=\sqrt{\Delta |p_{\theta}| /|p_{\perp}|^{2}}$, $\bar{\mu}_{\perp}=\sqrt{\Delta /|p_{\theta}|}$, and $\Delta$ is a constant equal to the 
minimum nonzero eigenvalue allowed in LQG for the area operator. This choice for $\bar{\mu}_j$ is motivated by consistency arguments and is usually called the \emph{improved dynamics} prescription \cite{aps}. The state dependence of  $\bar{\mu}_j$ makes the action of $\hat{\mathcal{N}}_{\bar\mu_{j}}$ on  $|p_j\rangle$ quite complicated. To simplify it,  it is better to relabel the basis states as $|\lambda,v\rangle$, with  $\lambda=\alpha^{-1/3}\text{sgn}(p_{\theta})\sqrt{|p_{\theta}|}$ and $v=\lambda w$, where $w=2\alpha^{-2/3}p_{\perp}$ and $\alpha=4\pi\gamma\hbar G\sqrt{\Delta}$. The label $v$ is proportional to the \emph{homogeneous} volume, $V_{\text{hom}}=(\alpha/2) v$. The operators $\hat{\mathcal{N}}_{\pm\bar{\mu}_{\theta}}$ and $\hat{\mathcal{N}}_{\pm\bar{\mu}_{\perp}}$ \emph{scale} the value of $\lambda$ and $w$, respectively, by a $v$-dependent factor, such that the label $v$ is simply \emph{shifted} by one unit \cite{awe}.

On the other hand, we describe the inhomogeneous sector employing the creation and annihilationlike variables chosen in \cite{cmmv,ccm}, since it has been proven to provide a unique satisfactory Fock quantization of the associated deparametrized model. The representations for both matter and gravitational inhomogeneities are given by identical Fock spaces, $\mathcal{F}^{f} (f=\xi,\varphi)$, obtained by promoting each pair of creation and annihilationlike variables to operators. An orthonormal basis is provided by the $n$-particle states. We call $n_m^{f}$ the occupation number of the field $f$ in the mode $m$. Finally, for the homogeneous matter field, $\phi$, a standard Schrödinger representation is used, with $\hat{p}_{\phi}=-i\hbar\partial_{\phi}$.

The representation of the momentum constraint in terms of the creation and annihilation operators is straightforward.
The imposition of this constraint leads to the  condition $\sum_{f}\sum_{m=1}^{\infty}m(n^f_{m}-n^{f}_{-m})=0$, \cite{hybrid5}. The  states that satisfy this condition form a proper Fock subspace, $\mathcal{F_{\text{p}}}$, of the inhomogeneous Hilbert space,  $\mathcal{F}^{\xi}\otimes
\mathcal{F}^{\varphi}$.

For the construction of the Hamiltonian constraint operator we follow the procedure of the vacuum case \cite{hybrid3}, so that the operator decouples the states of zero homogeneous volume and does not relate states with different sign of any of the labels
$\lambda$ and $v$. Consequently, we can remove from our theory the states
which are the analog of the classical singularity (those with $v=0$) and restrict the study
to the sector with e.g. strictly positive labels $\lambda$ and $v$. The Hamiltonian constraint operator reads
\begin{equation}
\widehat{\mathcal{C}}=-\dfrac{\widehat{\Theta}_{\perp}^{2}+\widehat{\Theta}_{\theta}\widehat{\Theta}_{\perp}+\widehat{\Theta}_{\perp}\widehat{\Theta}_{\theta}} {8\pi G \gamma^{2}}-\dfrac{\hbar^{2}}{2}\partial^{2}_{\phi}\!+ 2\pi\hbar\widehat{|p_{\theta}|}\widehat{H}_{0}+\widehat{\left[\dfrac{1}{|p_{\theta}|^{\frac{1}{4}}}\right]}^{2}\!\!\!\dfrac{\hbar\widehat{\Theta}_{\perp}^{2}}{4\pi \gamma^{2}}\widehat{\left[\dfrac{1}{|p_{\theta}|^{\frac{1}{4}}}\right]}^{2}\!\!\widehat{H}_{\text{int}}.
\end{equation}
Here we have introduced the quantum version of
${c_{j}p_{j}}$, given by
\begin{equation}
\widehat{\Theta}_{j}=i\pi \gamma  G \hbar \widehat{\sqrt{|v|}}\hspace*{-0.5mm}\left[\hspace*{-0.5mm}\left(\hspace*{-0.5mm}\hat{\mathcal{N}}_{ -2\bar{\mu}_{j}}\hspace*{-0.5mm}-\hat{\mathcal{N}}_{2\bar{\mu}_{j}}\hspace*{-0.5mm}\right)\widehat{ \text{sgn}(p_{j})}\hspace*{-0.5mm}+\widehat{{\text{sgn}}(p_{j})}\left(\hspace*{-0.5mm}\hat{\mathcal{ N}}_{-2\bar{\mu}_{j}}\hspace*{-0.5mm}-\hat{\mathcal{N}}_{2\bar{\mu}_{j}}\hspace*{-0.5mm} \right)\right]\widehat{\sqrt{|v|}},
\end{equation}
and the operator $\widehat{\left[1/|p_{\theta}|^{1/4}\right]}$, which is regularized as usual by taking the commutator of some suitable power of $\hat{p}_\theta$ with $\hat{\mathcal{N}}_{\bar{\mu}_{\theta}}$ \cite{hybrid3}. The operators $\widehat{H}_{0}$ and $\widehat{H}_{\text{int}}$ act nontrivially  only on the inhomogeneous sector of the kinematical Hilbert space. The action of $\widehat{H}
_{0}$ is diagonal on the $n$-particle states, while the action of $\widehat{H}_{\text{int}}$ is more involved and creates and annihilates pairs of particles \cite{hybrid5}. It is important to note that the inhomogeneities of both fields contribute to the constraints in exactly the same way.

The Hamiltonian constraint operator does not relate all states with different values of $\lambda$ and $v$, but \emph{superselects} different sectors. The superselection sectors in $v$ are semilattices of step four in $\mathbb{R}^{+}$ with a minimum value $\epsilon \in (0,4]$. The superselection sectors in $\lambda$ are more complicated. Given initial data $\lambda^{\ast}$ and $\epsilon$, the values of $\lambda$ in the corresponding sector are of the form $\lambda=\omega_{\epsilon}\lambda^{\ast}$, where $\omega_{\epsilon}$ runs over a certain (well known) countable set, that is dense in $\mathbb{R}^{+}$ (see \cite{hybrid3}).

The quantum Hamiltonian constraint leads to a difference equation in the variable $v$, and then we can regard it as an evolution equation in $v$. As in the vacuum Gowdy model \cite{hybrid3}, the solutions are completely determined by a dense set of data on the initial section $v=\epsilon$. This property allows us to characterize the physical Hilbert space as the Hilbert space of these initial data, whose inner product is determined by imposing reality conditions in a complete set of observables.

\section{Conclusions}
We have obtained a complete quantization of the LRS-Gowdy $T^{3}$ model with linearly polarized gravitational waves and a minimally coupled inhomogeneous massless scalar field as matter content using hybrid techniques in the framework of LQC. This hybrid quantization follows exactly the lines developed in the vacuum model, because matter and gravitational inhomogeneities can be treated in the very same way.

Owing to the polymeric quantization of the homogeneous gravitational sector, we have been able to eliminate the states analog to the classical cosmological singularity at the level of superselection. Furthermore, the standard quantization of both matter and gravitational inhomogeneities is recovered on physical states.

The inclusion of the massless scalar field not only allows us to study matter inhomogeneities in the framework of LQC by means of a hybrid quantization, but also leads to a model which is physically more interesting. In fact, the homogeneous sector of the model admits now flat Friedmann-Robertson-Walker (FRW) solutions. This hybrid LRS-Gowdy $T^{3}$ model provides a perfect scenario to study in further detail the quantum back-reaction of anisotropies and inhomogeneities on a loop-quantized flat FRW background, as well as the development of approximated methods in LQC aimed at extracting physical predictions from the theory.

\section*{Acknowledgements}
This work was supported by the Spanish MICINN/MINECO Projects No. FIS2008-06078-C03-03 and
No. FIS2011-30145-C03-02, and the Consolider-Ingenio Program CPAN No. CSD2007-00042. D. M-dB is
supported by CSIC and the European Social Fund under the Grant No. JAEPre-09-01796.

\end{document}